\documentclass[amsmath, amssymb, english, aps, prl, twocolumn,
floatfix, superscriptaddress, showpacs]{revtex4}
\usepackage[english]{babel}
\usepackage{graphicx}
\usepackage{verbatim}
\usepackage{natbib,doi,hyperref,url}



\def\expect#1{\mathinner{\langle{#1}\rangle}}

{\catcode`\|=\active
  \gdef\expect#1{\left<\mathcode`\|"8000\let|\bravert {#1}\right>}}
\def\bravert{\egroup\,\vrule\,\bgroup}

\begin{document}

\title{

Spectral Properties of Correlated Materials:
Local Vertex and Non-Local Two-Particle Correlations from 
Combined GW and Dynamical Mean Field Theory

}

\author{Thomas Ayral}
\affiliation{Theoretical Physics, ETH Zurich, 8093 Z\"urich, Switzerland}
\affiliation{Centre de Physique Th\'eorique, Ecole Polytechnique, CNRS UMR7644,
91128 Palaiseau Cedex, France}
\affiliation{Institut de Physique Th\'eorique (IPhT), CEA, CNRS, URA 2306, 91191 Gif-sur-Yvette, France
}

\author{Philipp Werner}
\affiliation{Department of Physics, University of Fribourg, 1700 Fribourg, Switzerland}
\affiliation{Theoretical Physics, ETH Zurich, 8093 Z\"urich, Switzerland}

\author{Silke Biermann}
\affiliation{Centre de Physique Th\'eorique, Ecole Polytechnique, CNRS UMR7644,
91128 Palaiseau Cedex, France}
\affiliation{Japan Science and Technology Agency, CREST, Kawaguchi
  332-0012, Japan}

\begin{abstract}

We present a fully self-consistent combined GW and dynamical mean field
(GW+DMFT)
study of the spectral properties of the extended two-dimensional Hubbard 
model. The inclusion of the local dynamical vertex
stemming from the DMFT self-energy and polarization
is shown to cure the problems of self-consistent GW
in the description of spectral properties. 
We calculate the momentum-resolved spectral functions, the two-particle 
polarization and electron loss spectra, and 
show that the inclusion of GW in  extended DMFT leads to a narrowing 
of the quasi-particle width and more pronounced 
Hubbard bands in the metallic regime as one approaches the 
charge-ordering transition. 
Finally, the momentum-dependence introduced by GW into the extended 
DMFT description of collective modes is found to 
affect their shape, giving rise to dispersive plasmon-like long-wavelength 
and stripe modes.

\end{abstract}

%check PACS !!!

\maketitle

Modern spectroscopic techniques are able to measure
one- and two-particle spectra of condensed matter 
systems with remarkable precision, characterizing
not only quasi-particle excitations but unveiling
also satellite structures. Examples
include Hubbard bands in photoemission spectroscopy,
stemming from the atomic-like behavior of the electrons 
in partially filled narrow $d$- or $f$-shells
\cite{Imada_1998}, 
or collective excitations such as plasmonic features.
Addressing such effects requires
an accurate description of one- and two-particle
spectral functions within the framework of many-body 
theory.  The quantitative prediction of satellite
features has even been used as a quality marker for
many-body techniques. The failure of self-consistent
perturbation theory in the screened Coulomb interaction,
the self-consistent GW approximation,
to describe plasmon satellites in the electron gas
for example has provided arguments in favor of a non-self-consistent
(``one-shot GW'') treatment \cite{vonBarth_1996,Holm_1998}. 
For real solids,
few fully self-consistent calculations are available 
\cite{Ku_2002, Kutepov_2010}, and no consensus concerning the 
virtues of self-consistency has been reached so far.
A popular scheme, dubbed quasi-particle self-consistent
GW (QPSC-GW) \cite{Kotani_2007} yields reasonable estimates 
both for total energies and spectra. 
Yet, most of the calculations within this scheme
were applied to semiconductors, and applications to correlated 
metals only start to appear \cite{Kutepov_2011}.
The inclusion of an appropriate vertex correction
is expected to resolve the ambiguities around the
self-consistency question, and it has been in particular
proposed that a combined GW and dynamical mean field
scheme \cite{Biermann_2003} would enable 
self-consistent calculations even for spectral properties.
Early pioneering calculations on a three-dimensional 
extended Hubbard model \cite{Sun_2002, Sun_2004}
have benchmarked several flavors of combined schemes along these lines.
However, the numerical difficulty of solving the DMFT
equations with frequency-dependent interactions
has until recently prevented the direct investigation of spectral properties.

Implementing the combined GW+DMFT scheme in a fully 
self-consistent manner for the two-dimensional
extended Hubbard model, 
we here demonstrate that this technique indeed successfully 
overcomes the deficiencies of GW. 
The implicit inclusion of a non-perturbative local vertex  
enables fully self-consistent calculations 
for spectral properties. In the correlated metal regime, 
the GW+DMFT self-energy encodes both, 
band renormalization effects and Hubbard satellite features. 
As expected from the physical ingredients, the theory also 
describes the Mott insulating state for strong local 
Coulomb interaction, which is inaccessible in GW alone, 
as well as the charge-ordered state driven by intersite 
interactions, absent from standard DMFT.
In addition, dynamical screening effects give rise to plasmonic 
features in the local spectral function. 
While the local spectral functions in the intermediate
to strong correlation regime are little affected by the
non-local self-energy contributions stemming from the
GW approximation, 
a substantial $k$-dependent modulation of the peak widths
is observed in the momentum-resolved spectral functions.
We analyze momentum-resolved 
two-particle spectra and show that the self-consistent combination 
of GW and EDMFT strongly affects the shape of collective modes, 
giving rise to dispersive plasmon-like long-wavelength modes 
and stripe modes.

We consider the half-filled extended Hubbard model on 
a two-dimensional square lattice 
\begin{align}
H=-t\sum_{i\ne j,\sigma} c_{i\sigma}^{\dagger}c_{j\sigma}%-\mu\sum_{i}n_{i}
+\sum_{i}(Un_{i\uparrow}n_{i\downarrow}-\mu n_i)+ \frac{V}{2} \sum_{i\ne j}n_{i}n_{j},\nonumber
\label{eq:U-V_model}
\end{align}
where $c_{i\sigma}$ and $c_{i\sigma}^{\dagger}$ denote the annihilation
and creation operators of a particle of spin $\sigma=\uparrow,\downarrow$
at the lattice site $i$, $n_{i\sigma}=c_{i\sigma}^{\dagger}c_{i\sigma}$, and
$n_{i}=n_{i\uparrow}+n_{i\downarrow}$. $\sum_{i\ne j}$ is the sum over all nearest-neighbor sites, $t>0$ is the hopping amplitude between two neighboring
sites, $\mu$ is the chemical potential, $U$ the on-site repulsion
between electrons of opposite spin and $V$ the repulsion between
two electrons on neighboring sites. The Fourier-transformed bare 
interaction term thus reads $v_{k} = U+2V\left(\cos(k_{x})+\cos(k_{y})\right)$. 
All energies are given in units of the half-bandwidth $D=4t$.
We show results for inverse temperature $\beta D=100$,
restricting our study to the paramagnetic phase.

The GW+DMFT approach is derivable from a free energy functional 
\cite{biermann-gwdmft-arxiv0401653}. 
The Legendre transform of the free energy with respect to 
the Green's function $G$ and the screened interaction $W$ can
be expressed as a sum of the Hartree-Fock part and a 
Luttinger-Ward-like correlation functional $\Psi[G,W]$, which 
sums up all skeleton diagrams built from $G$ and $W$ \cite{Almbladh_1999}.
The  GW+DMFT scheme consists in approximating $\Psi$ as 
$\Psi\approx\Psi^{\mathrm{EDMFT}}[G_{ii},W_{ii}]
+\Psi^{\mathrm{GW}}_{\mathrm{\mathrm{nonloc}}}[G_{ij},W_{ij}]$, where 
the first term is calculated from a (dynamical) impurity problem
as in {\it extended} dynamical mean field theory (EDMFT) 
\cite{Sengupta_1995, Si_1996, Kajueter_thesis} and
the second term is the 
non-local ($i\ne j$) part of the GW functional 
$\Psi^{\mathrm{GW}}_{\mathrm{\mathrm{nonloc}}}[G_{ij},W_{ij}] = 
\sum_{i \ne j} G_{ij}W_{ij}G_{ji}$.

The GW+DMFT scheme self-consistently constructs the Green's function 
$G$ and the screened interaction $W$ of the system as a stationary 
point of the free energy functional.
The self-energy $\Sigma$ and polarization $P$ are formally obtained 
by functional differentiation of $\Psi$ with respect to $G$ and $W$, 
leading to the expressions 
$\Sigma(k,i\omega)=\Sigma_{\mathrm{imp}}(i\omega_n)+\Sigma^{\mathrm{GW}}_{\mathrm{nonloc}}(k,i\omega)$ 
and 
$P(k,i\nu_n) = P_{\mathrm{imp}}(i\nu_n)+P_{\mathrm{nonloc}}^{\mathrm{GW}}(k,i\nu_n)$ 
($\omega_n$ and $\nu_n$ are fermionic and bosonic Matsubara frequencies, 
respectively). This endows GW+DMFT with conserving 
properties \cite{Baym_1962}. 
The momentum-dependent $G$ and $W$ are then calculated from the 
one- and two-particle Dyson
equations and used as inputs for a GW calculation, yielding 
$\Sigma^{\mathrm{GW}}$ and $P^{\mathrm{GW}}$. Meanwhile, their local parts are extracted 
to compute the local Weiss fields $\mathcal{G}$ and $\mathcal{U}$:
$\mathcal{G}^{-1}(i\omega_n)  =  G_\text{loc}^{-1}(i\omega_n)+\Sigma_\mathrm{imp}(i\omega_n)$ and
$\mathcal{U}^{-1}(i\nu_n)  =  W_\text{loc}^{-1}(i\nu_n)+P_\mathrm{imp}(i\nu_n)$.
These, in turn, are used as inputs to a \emph{dynamical} impurity model, which we solve using a continuous-time Monte Carlo algorithm \cite{Werner_2007, Werner_2010} to obtain
updated local self-energies. The whole scheme is iterated until convergence. 
The calculations have been performed 
on a $64\times64$ momentum grid, 
while the analytical continuation of the imaginary-time data has 
been performed using the Maximum Entropy method \cite{Jarrell_1996}.
We monitor the following quantities: 
(i) the local spectral function 
$A_\text{loc}(\omega) = - \frac{1}{\pi} \mathrm {Im}  G_\text{loc}(\omega)$, 
(ii) the local and non-local self-energy, 
(iii) the local and non-local polarization, 
(iv) the electron energy-loss spectrum (EELS) $\mathrm{Im}\left[-\epsilon(k,\omega)^{-1}\right]$
(where $\epsilon$ is the dielectric function, $\epsilon(k,\omega) = 1-v_{k}P(k,\omega)$).

%%%%%%%%%%%%%%%%%%%%%%%%%%%%%%%%%%%%%%%%%%%%%%%%%%%%%

\begin{figure}
   \includegraphics[width=\linewidth,keepaspectratio]{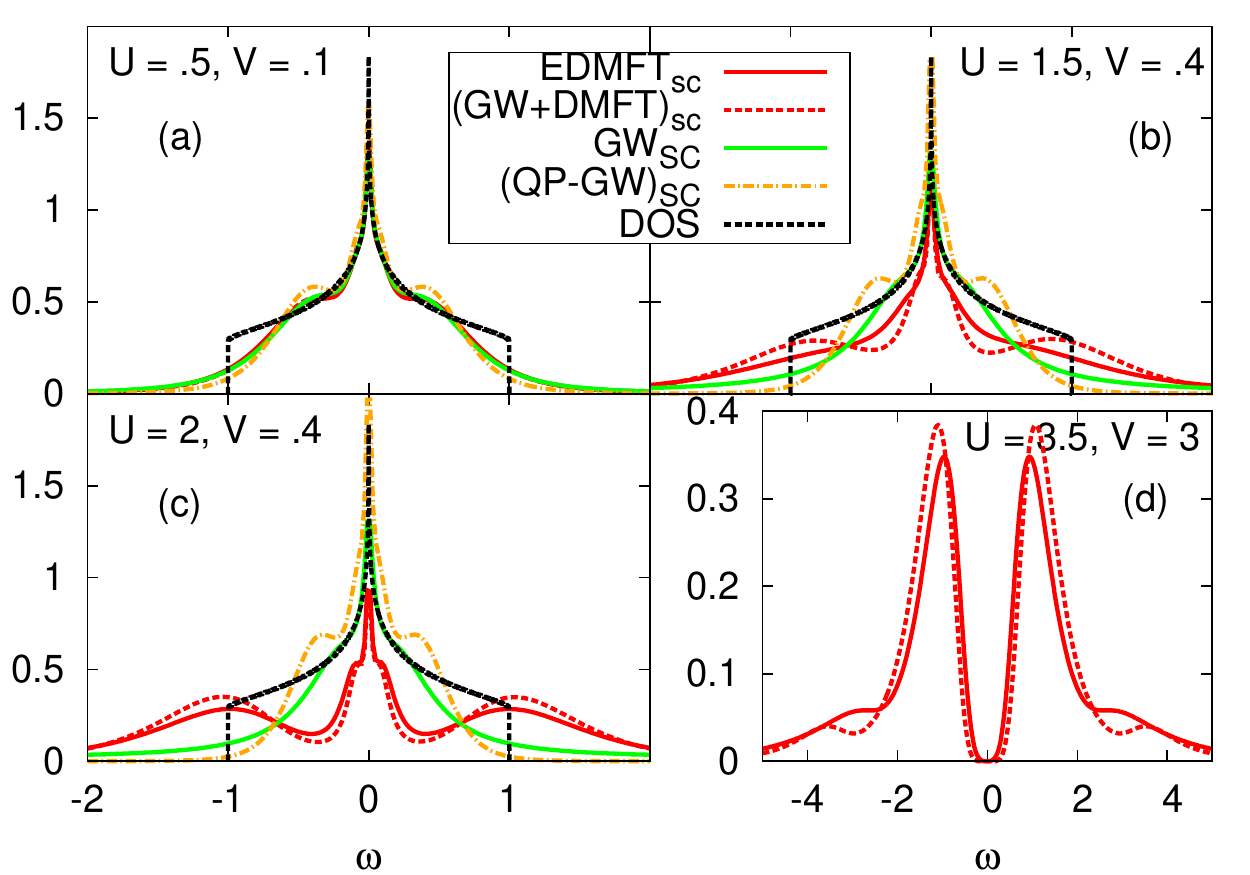}
   \caption{\label{fig1} 
Spectral function $A_\text{loc}(\omega)$ obtained within four different self-consistent schemes (see text). 
   }
 \end{figure}

Within extended DMFT and GW+DMFT, in the absence of intersite repulsion, the Mott transition takes place at $U_c\approx 2.5$. This value is slightly modified by intersite repulsions $V<V_c=0.8$. At $V_c$ a transition to a charge-ordered phase occurs.
In the following we study the local spectral properties in the metallic
phase with weak ($U=0.5$, $V=0.1$) and intermediate ($U=1.5$-$2$, $V=0.4$)
interactions as well as in the Mott insulator at $U=3.5$ and $V=3$.
Figure~\ref{fig1} shows the local spectral function $A_\text{loc}(\omega)$  
obtained within 
(i) (self-consistent) EDMFT, (ii) self-consistent GW+DMFT,
(iii) self-consistent GW and (iv) quasi-particle self-consistent GW 
(QPSC-GW). The latter scheme was implemented by
computing the lattice Green's function from the GW self-energy 
via $G(k,i\omega_n)^{-1} = i \omega_n - Z_k (\epsilon_k - \mathrm{Re}\Sigma_{\mathrm{GW}}(k,i\omega_0))$, 
where 
$Z_k\approx(1- \mathrm{Im}\Sigma_{\mathrm{GW}}(k,i\omega_0)/\omega_0)^{-1}$ 
is the quasi-particle weight as estimated from the value
of the self-energy at the first Matsubara frequency.  
For small interactions 
(panel 1a), correlation effects are
negligible, and the four schemes result in 
indistinguishable spectra within the numerical accuracy.
As the local
interaction $U$ becomes significant (panel 1b)
deviations start to appear, with EDMFT and GW+DMFT exhibiting
stronger correlation effects. 
Upon increasing local interactions (panel 1c), the quasi-particle
renormalization becomes stronger, the width of the
coherent central peak shrinks, and the corresponding
spectral weight is transferred to higher energies.
This -- physically expected -- behavior 
is realized by the EDMFT and GW+DMFT spectra, which exhibit
higher-energy structures at $\omega\approx \pm U/2$ already for 
$U=1.5$. The Hubbard bands gain spectral weight as $U$ 
increases further. Interestingly, they are more pronounced within
GW+DMFT than within EDMFT. Finally, at $U=3.5$, a Mott gap
has opened, and EDMFT and GW+DMFT spectra are similar.
In addition to the two Hubbard bands, the EDMFT and GW+DMFT
spectra display two symmetric high-energy satellites, whose 
spectral weight depends on the intersite interaction $V$. 
The QPSC-GW spectra display only a weak renormalization of 
the bandwidth as $U$ increases
from the weak to the strong coupling limit, and at all
correlation levels the spectra remain metallic. 
The same is true within the self-consistent GW method, 
although with increasing correlations some spectral weight 
is shifted to higher frequencies, albeit in a featureless way. 

These observations show that both self-consistent GW approaches yield 
a correct result only in the weak-correlation regime. As correlations increase, 
GW fails to describe the shift of spectral weight to high-energy 
incoherent bands, present in DMFT. We note that in the local GW+DMFT  
spectra the Hubbard bands are enhanced compared to the EDMFT or GW spectra. 
This effect can be ascribed to the self-consistency, which allows the local 
quantities to re-adjust to the non-local self-energies $\Sigma_{\mathrm{GW}}$ and 
$P_{\mathrm{GW}}$.

Another salient characteristic of EDMFT and GW+DMFT spectra is the presence 
of additional high-energy satellites in the Mott phase. These directly 
reflect the frequency-dependence of the local interactions 
$\mathcal{U}(\omega)$ induced by the nearest-neighbor repulsion term $V$. 
As demonstrated in Ref.~\cite{Casula_2012}, a pole in $\mathcal{U}(\omega)$ (such as a 
plasmon pole) leads to multiple satellites in the local spectral function. 
In our case, the local interactions $\mathcal{U}(\omega)$ are characterized 
by a broad continuum of poles centered at some energy $\omega_d$, resulting 
in only two symmetric satellites in the Mott spectra. In the metallic phase, 
these satellites are present, but they are broad and merged with the 
Hubbard bands, making them hardly distinguishable.

\begin{figure}
   \includegraphics[width=\linewidth,keepaspectratio]{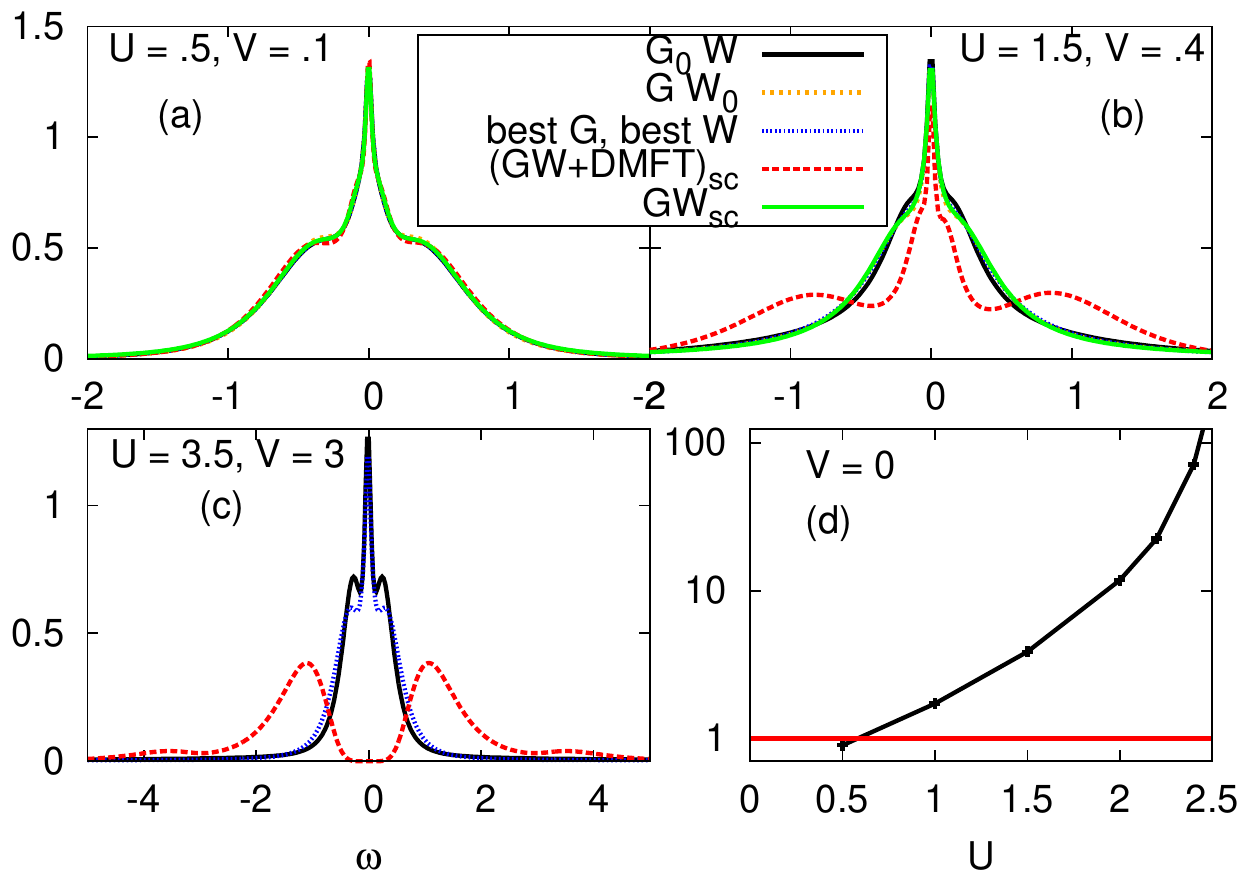}
   \caption{\label{fig2} 
Spectral function $A_\text{loc}(\omega)$ obtained using 3 different one-shot GW schemes (see text). Lower right panel: vertex estimate $\Lambda(\omega=0)$ as a function of $U$.  }
 \end{figure}

\begin{figure}
\includegraphics[width=\linewidth,keepaspectratio]{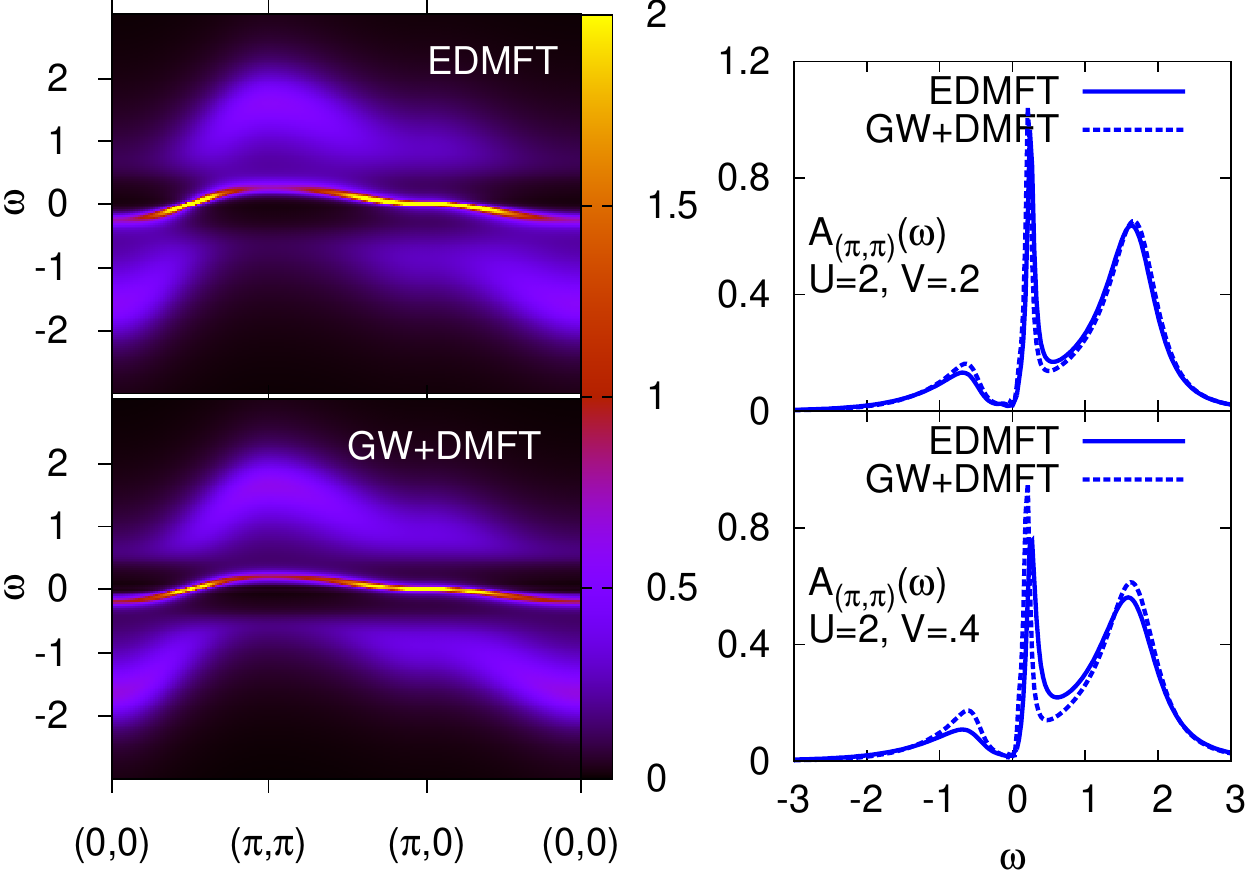}
\caption{ 
Left panels: momentum-resolved spectral function 
$A(k,\omega)$ at $U=2$, $V=0.4$ within EDMFT (top)  and GW+DMFT (bottom).
Right panels: $A(k,\omega)$ at $k=(\pi,\pi)$ for $U=2$, $V=0.2$ (top) and $U=2$, $V=0.4$ (bottom) within both schemes.
\label{fig3}
}  
\end{figure}

The failure of both self-consistent GW schemes to capture Hubbard bands 
or high-energy satellites is consistent with the well-known observation 
that self-consistency in GW for the homogeneous electron gas results in 
a smearing out and displacement of high energy satellite features 
\cite{vonBarth_1996}.
In light of this observation, most modern GW schemes therefore adopt a ``best-G-best-W'' strategy, rather than aiming at full self-consistency.
Figure 2 illustrates the virtues and limitations of this strategy
by displaying the spectra obtained in different one-shot GW schemes: 
(i) in ``$G_0 W$", the non-interacting Green's function $G_0$ and the 
converged GW+DMFT $W$ are taken as inputs to a one-shot GW calculation, 
(ii) ``$G W_0$" takes the converged GW+DMFT $G$ and $W_0=v(1-v G_0 G_0)^{-1}$ 
within the random phase approximation as inputs and 
(iii) ``best G, best W" takes the converged GW+DMFT $G$ and $W$ as inputs. 
At all correlation levels ($U=0.5$ to $U=3.5$), these three
GW schemes 
produce results very similar to self-consistent GW. 
In particular, they remain metallic. Even the ``best G, best W" scheme in 
the Mott phase ($U=3.5$) yields a metallic self-energy, despite the 
Mott-like character of the input $G$ and $W$. This phenomenon 
is due to the lack of Hedin's three-legged
 vertex $\Lambda$ in GW schemes, as shown in 
the lower-right panel of Figure 2. There, an estimate of the local part of $\Lambda$ is computed from EDMFT results at $V=0$. 
Remembering that, schematically, the irreducible vertex function 
$\Lambda$ appears in the self-energy as  $\Sigma = G W \Lambda$ 
\cite{Hedin_1965}, a rough estimate 
-- neglecting the true frequency structure  --
is computed 
as follows: one computes an effectively vertex-corrected screened 
interaction 
$\tilde{W}(\tau) = \Sigma_{\mathrm{imp}}(\tau)/G_{\mathrm{imp}}(\tau)$ from EDMFT, 
then Fourier-transforms it to $\tilde{W}(i\nu_n)$; finally, the static 
vertex estimate is obtained as $\Lambda(0)\approx \tilde{W}(i\nu_0)/W_\text{loc}(i\nu_0)$.
Crude as it is (the full vertex depends on two independent frequencies), 
this estimate nonetheless clearly demonstrates the role of vertex corrections 
for the Mott transition: from unity in the weakly correlated regime, it 
increases with $U$ until it diverges at the Mott transition. 
This indicates that within the language of Hedin's equations, the divergence 
of the local vertex is the driving force of the Mott phenomenon, making any vertex-less approximation unfit to capture it.

\begin{figure}
\includegraphics[width=\linewidth,keepaspectratio]{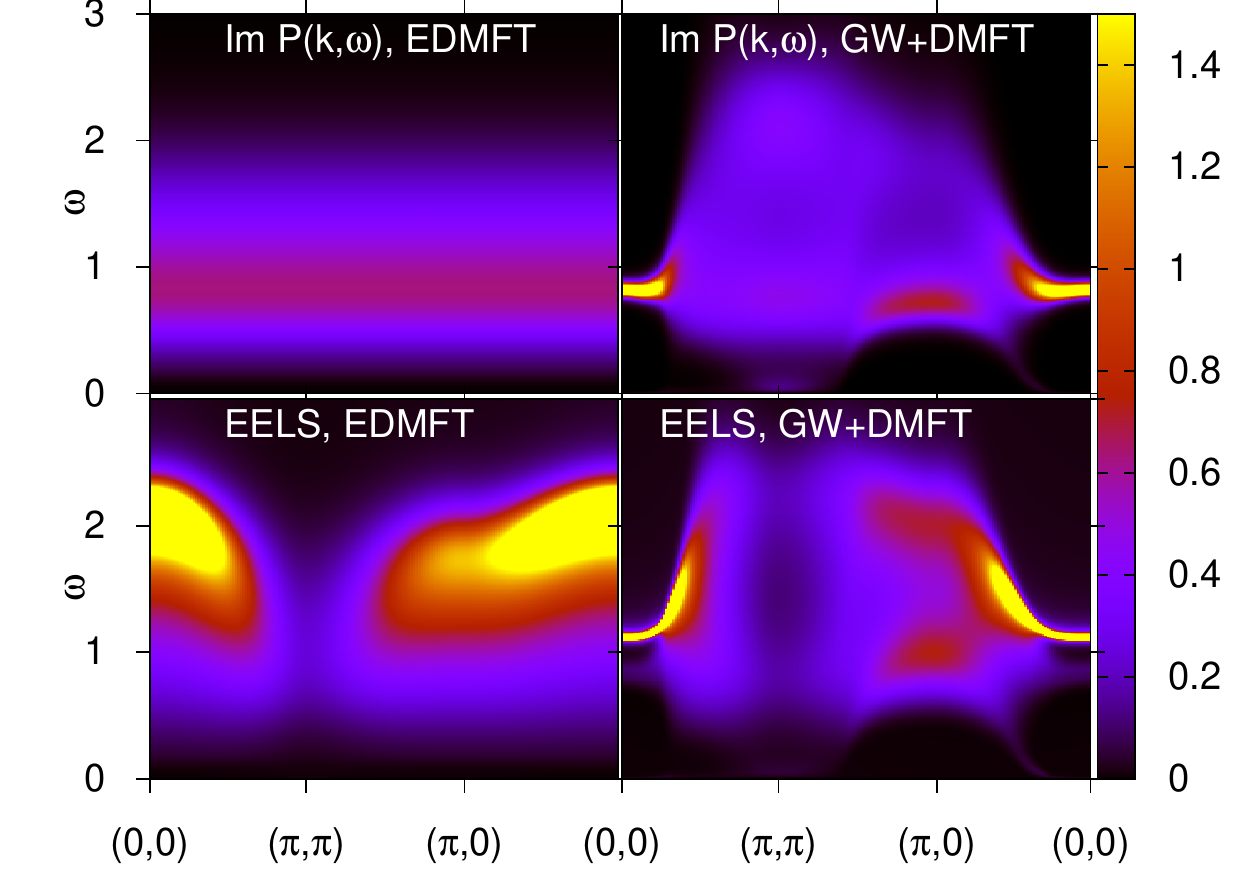}
\caption{ 
$U=2, V=0.4.$ 
Upper panels: $\mathrm{Im} P(k,\omega)$ within EDMFT (left) and GW+DMFT (right). 
Lower panels: $\mathrm{Im}\epsilon^{-1}(k,\omega)$ within EDMFT (left) and GW+DMFT (right).
\label{fig4}
 }
 \end{figure}

The effect of the non-local GW contributions on EDMFT are  
illustrated by the momentum-resolved spectral functions, displayed 
in Figure~\ref{fig3}. Generally, the dispersion of the Hubbard bands
follows the dispersion of the quasi-particle peak, within both 
schemes.
In the presence of a strong intersite interaction, the non-local
self-energy and polarization lead however to an additional $k$-dependent
modulation of the linewidth and weights. 
This is illustrated in the lower panels of
Fig.~\ref{fig3} for the $(\pi, \pi)$ point, where a pronounced sharpening
of the quasi-particle peak is observed along with enhanced weight
of the Hubbard bands.

We now turn to a study of two-particle quantities.
Figure \ref{fig4} shows the momentum-resolved imaginary part of the 
polarization and the electron energy loss (EELS) spectrum 
$\mathrm{Im}\left[-\epsilon^{-1}(k,\omega)\right]$ in the metallic regime. 
Within EDMFT, the polarization displays a broad mode which reflects 
the particle-hole excitations of the system. They are centered at 
$U/2$, reflecting the emergence of the Hubbard bands and the 
corresponding excitations between Hubbard bands and the 
quasi-particle peak.
In contrast, the polarization spectrum within GW+DMFT is dispersive. 
While displaying 
sharper features close to the $\Gamma\equiv(0,0)$ point, it 
captures particle-hole excitations due to Fermi-surface nesting at
wave-vector $(\pi,\pi)$, as well as the zero-sound mode at 
long wavelengths and low energies.
The EELS spectrum contains the particle-hole excitations 
(poles of the polarization) of the system and its collective 
modes, which correspond to the solutions of 
$\mathrm{Re} P(k,\omega) = 1/v_k$. These collective modes are 
damped out close to particle-hole excitations (when 
$\mathrm{Im} P(k,\omega)$ is large). This analogue of the 
free-electron-gas Landau damping occurs at the 
$(\pi,\pi)$ point in EDMFT and GW+DMFT. It can be directly 
ascribed to the nearest-neighbor repulsion, which induces 
scattering along this direction. The energy and lifetime of 
this collective excitation differs from EDMFT to GW+DMFT. 
In GW+DMFT it is lower in energy, more dispersive and with a larger lifetime.
In GW+DMFT, two modes are visible above the $(\pi,0)$ point, 
indicating the existence of two stripe modes at energies 
$\omega = 1$ and $\omega=2$ corresponding to
stripe-like modulations, where the sign of the density fluctuation
varies from row to row in the $x$-direction.
For obvious reasons, they are not captured by EDMFT.
These two-particle excitations are directly related to the 
screening in the system as the screened interaction, $W$, is given 
by $W(k,\omega) = \epsilon^{-1}(k,\omega) v_k $. 
In particular, they explain the retardation effects at play in 
the local interactions $\mathcal{U}(\omega)$ and the corresponding 
satellites in the local spectra.

In conclusion, we have demonstrated how the ambiguities on 
the optimal degree of self-consistency in many-body perturbation
theory are resolved by including a non-perturbative local vertex
in the calculation. Based on a fully self-consistent implementation
of the combined GW+DMFT scheme, we have analyzed one- and 
two-particle satellite features in correlated materials.
While we confirm the well-known ``washing out'' of satellite
features in self-consistent GW calculations, self-consistent
GW+DMFT does not suffer from this deficiency.
Plasma- and zero-sound-like oscillations involving itinerant carriers 
as in the electron gas survive only in the regime of small local 
Coulomb interactions, but are quickly suppressed in the correlated
metal. In this regime, excitations related to the 
creation of doublons become dominant.
While local spectral functions are little affected by non-local
contributions in a wide range of parameters, the momentum-resolved
spectra display a $k$-dependent modulation of the width of the
peaks; the momentum-dependence introduced by the GW part becomes
truly crucial when assessing dispersions of two-particle spectral
properties, differentiating in particular the nature of the collective
modes in the $(0,0)$, $(0,\pi)$, and $(\pi, \pi)$ directions.
Our findings have implications for the nature of satellite
features in correlated materials. In particular, it becomes
obvious that electron-gas-like plasmons in materials stem 
dominantly from the charge contained in completely {\it filled} 
shells (that is from multi-orbital effects), while partially 
filled shells give rise
to doublon excitations of the kind we describe.
The interplay of local correlations and charge ordering phenomena
and their intriguing wave-vector dependence call 
for an extension of our study to 
realistic two-dimensional
systems: recent experimental findings of charge ordering in
cuprates \cite{Wu_2011}, 
cobaltates \cite{Mukhamedshin_2005} 
or in systems of adatoms on surfaces \cite{Hansmann_2012}
are prominent examples.

\acknowledgments{
We acknowledge useful discussions with 
F. Aryasetiawan, M. Casula, A. Georges, P. Hansmann,
M. Imada, M. Katsnelson, A. Millis, and T. Miyake.
This work was supported by the French ANR under project SURMOTT,
GENCI/IDRIS Orsay under project 1393, by DFG FOR 1346 and by the 
Swiss National Science Foundation (grant PP0022-118866).
Most of the calculations have been performed on the Brutus 
cluster at ETH Zurich, using a code based on ALPS \cite{ALPS}.  
We thank L.~Boehnke for allowing us to use his MaxEnt code.
}

%\bibliography{refs_thomas}

\end{document}